# CLIC DRIVE BEAM AND LHC BASED FEL-NUCLEUS COLLIDER

H. Braun, R. Corsini, CERN, Geneva, Switzerland
S. Sultansoy, Gazi University, 06500, Ankara, Turkey
O. Yavas, Ankara University, 06100, Ankara, Turkey

*Abstract*
The feasibility of a CLIC-LHC based FEL-nucleus collider is investigated. It is shown that the proposed scheme satisfies all requirements of an ideal photon source for the Nuclear Resonance Fluorescence method. The physics potential of the proposed collider is illustrated for a beam of Pb nuclei.

## 1. INTRODUCTION

The investigation of collective excitations, especially electric and magnetic dipole vibrations of nuclei plays an important role in determination of nuclear structure. The Nuclear Resonant Fluorescence (NRF) is the most efficient method to study these excitations [1]. However, the capacity of traditional NRF experiments for study of higher line excitations are limited by several factors [2].

Recently, a new method, namely FEL-Nucleus Collider, has been proposed [3,4] which satisfy all requirements on ideal photon sources [1]: high spectral intensity, good monochromaticity, tunability in a broad energy range and high degree of linear polarization. For these reasons, FEL-nucleus colliders will be powerful tools for experimental determination of energy levels, decay widths, spin and parity of excited nuclei. The physics search potential of FEL-nucleus collider was demonstrated for the $^{154}$Sm nucleus excitations in [5], where the TTF FEL is considered as the source of photon beam. The advantages of a FEL-RHIC collider with respect to traditional NRF methods are demonstrated in [6] for some excitation levels of $^{232}$Th nucleus.

In this paper, we investigated the feasibility and physics search potential of CLIC-LHC based FEL-Nucleus collider.

## 2. NUCLEAR SPECTROSCOPY AND PHOTON SOURCES

The data on electric and magnetic dipole excitations have revealed exciting new insights into nuclear structure. For example, low-lying dipole excitations (so-called M1 scissors mode) and dipole excitations near particle threshold in transition and deformed even-even nuclei are of considerable interest in modern nuclear structure physics. Both M1 and E1 excitations were studied systematically to test nuclear models. Unfortunately, spin and parity of a lot of excitations are not determined with existing experimental facilities.

There are several methods to produce photons for low energy photon scattering experiments. An ideal photon source for such experiments should have the following characteristics [1]:

- High spectral intensity $I = N_\gamma$ / eVs (number of photons per energy bin and second).
- Good monochromaticity ($\Delta E_\gamma / E_\gamma$).
- Tunability in a broad energy range.
- High degree of linear polarization ($P_\gamma \approx 100\%$).

Unfortunately up to now there are no such ideal sources available fulfilling all these requirements. Therefore, only with the advent of a new experimental facility with improved characteristics it will be possible to investigate in detail the fine structure of the magnetic and electric response. FEL-nucleus colliders will have all the required characteristics due to the use of a FEL beam as photon source.

Free electron lasers can produce a photon beam in a wide region of the electromagnetic spectrum from infra-red to hard X-ray using relativistic electron beams with small emittance and undulator magnets [7-9]. In the SASE FEL, a high quality beam of relativistic electrons is injected into an undulator, i.e. a straight set-up of magnetic dipoles with equal field strength and alternated polarities [8].

## 3. FEL-NUCLEUS COLLIDERS

The accelerated fully ionized nuclei will "see" the few keV energy FEL photons as a laser beam with MeV energy [3, 4]. A schematic view of FEL-Nucleus collider is given in Figure 1. Due to good monochromaticity ($\Delta E_\gamma / E_\gamma \approx \rho < 10^{-3}$-$10^{-4}$) with a typical obtainable number of photons of the order of $10^{13}\gamma$/bunch, tunability and better polarization ($P\gamma \approx 100\%$) of the FEL beam, FEL-nucleus colliders can be successfully used to investigate nuclear excitations with low multi polarity in a wide energy region. The needed energy of FEL photons can be expressed as:

$$\omega_0 = \frac{E^*}{2\gamma_A} = \frac{A}{Z}\frac{E^*}{2\gamma_p} \qquad (1)$$

where $E^*$ is the energy of the corresponding excited level, A and Z are the atomic and mass numbers of the nucleus, $\gamma_A$ and $\gamma_p$ are the Lorentz factors of the nucleus and proton, respectively. The luminosity of the FEL-nucleus collider can be expressed as,

$$L = \frac{n_\gamma n_A}{4\pi\sigma_x\sigma_y} n_b f_{rep} \qquad (2)$$

where $n_\gamma$ and $n_A$ are the number of particles in FEL and nucleus bunches, respectively, $\sigma_x$ and $\sigma_y$ are the transverse beam sizes, $n_b$ is the number of bunches per FEL pulse, $f_{rep}$ is pulse frequency.

Since the width $\Gamma$ of the excited states are smaller than the energy spread of colliding beams, the approximate value of the averaged cross sections has been found to be

$$\sigma_{ave} \approx \sigma_{res} \frac{\Gamma}{\Delta E_\gamma} \qquad (4)$$

where $\sigma_{res}$ is given by the well-known Breit-Wigner formula and $\Delta E_\gamma$ is the energy spread of the FEL beam in the nucleus rest frame; $\Delta E_\gamma/E_\gamma \leq 10^{-4}$, $E_\gamma \approx E_{exc.}$ The number of the produced excited levels is given by $R = L\sigma_{ave}$.

## 4. CLIC DRIVE BEAM BASED FEL-LHC COLLIDER

It is well known that energy range of electromagnetic low multi-polarity excitations is 2-20 MeV. The corresponding FEL energies can be obtained from Eq. (1). For example, we obtain a wavelength range from 0.4 nm to 4 nm (0.34-3.4 keV photon energy) for Pb excitations, taking into account that $\gamma_{Pb} \approx 2941$ for LHC [10, 11]. To obtain a FEL with these energies we need $E_e \approx 1.2 - 4$ GeV for $\lambda_u = 2.5$ cm and $B_u = 0.43$ T, where $\lambda_u$ is a distance between two identical poles of the undulator and $B_u$ is the rms field.

In [3-6], a FEL driven by a superconducting linac has been considered. In the case of LHC, and if CLIC is eventually built at CERN [12], it would be interesting to explore the possibility of using CLIC components to accelerate the FEL electron beam.

Potential advantages of a CLIC-based FEL would be the (partial) use of existing equipment, and the possibility of reaching a high acceleration gradient, resulting in a shorter linac. On the other hand, as pointed out in previous work [13], the use of high frequency RF for acceleration is disadvantageous in terms of beam energy spread, which is required to be small by FEL physics.

A more promising option, which we will explore in the following, is to use a fraction of the CLIC drive beam to power a high-frequency, normal-conducting linac dedicated to FEL operation. A natural choice of the linac frequency would be 15 GHz, which corresponds to the drive beam bunch repetition frequency (in CLIC the power is produced at the 2$^{nd}$ harmonic of the drive beam, 30 GHz). The linac will have a lower gradient and/or will be less efficient than the CLIC main linac, but the energy spread can be made smaller, thus satisfying FEL requirements (typically $\sigma_p < 10^{-3}$).

The LHC ion beam is composed of bunches with rms duration of 250 ps, separated by 100 ns. In a CLIC-based FEL, while the repetition rate is of the order of 200 Hz, the bunch spacing is linked to the time structure of the RF pulses that can be produced by the drive beam. Using only a drive beam pulse per cycle (~ 2 % of the total CLIC power) to accelerate a single electron bunch ($f_{rep}$ = 150 Hz, $n_b$ = 1) and assuming $\sigma_x = \sigma_y = 30$ μm, $n_A = 7 \cdot 10^7$ and $n_\gamma = 2.5 \cdot 10^{13}$ we can obtain a "minimum" value for the luminosity: $L = 2.3 \cdot 10^{27}$ cm$^{-2}$s$^{-1}$.

Although the luminosity thus obtained is not completely negligible, the situation is far from optimum. Some improvements can be expected by a somewhat better matching of the nucleus beam and the gamma beam sizes (16 μm and 30 μm, respectively); but the highest gain can be obtained by a more efficient use of the drive beam. Possibly the best option is to use part of the drive beam pulse train as it comes out of the delay loop (see Fig. 2).

Assuming the still tentative new CLIC parameter list found in [14], one see that at this stage a drive beam is composed by 70 ns pulses, spaced by 140 ns. Assuming for instance the use of an alternative, slightly shorter delay loop this can be brought easily to 50 ns long pulses, spaced by 100 ns. Each pulse can then be used to provide an RF pulse useful to accelerate an electron pulse. The total number of pulses available depends on how large a fraction of additional power one is willing to use. As an example, we consider using 10 μs, i.e., 10 % of the total CLIC drive beam pulse, which will provide $n_b$ = 100, increasing the luminosity by the same factor. Detailed consideration show that $L = 2.5 \cdot 10^{29} cm^{-2} s^{-1}$ at 4 nm wavelength.

## 5 PHYSICS SEARCH POTENTIAL

Decay widths, spin and parity of a lot of excited levels are not determined by NRF methods. A FEL-Nucleus collider will give opportunity to measure these quantities. For example, due to high statistics, unknown decay widths can be determined using known ones [3]. The spin of excited nucleus can be determined using the angular distribution of emitted photons [5]. Determination of the parity is considered in [6].

### 5.1 Pb example

Main characteristics of the $^{208}$Pb excitations with low spin (J≤2), observed in $(\gamma, \gamma')$ reactions are presented in the Table 2 [15]. Let us mention that there are a lot of other (J≤2) levels of the $^{208}$Pb which are observed in different reactions, such as (e,e$'$) etc. These levels are not observed by traditional NRF methods, but most of them will be seen at FEL-nucleus collider.

The needed FEL energies ($\omega_{FEL}$), resonant and average cross-sections and event numbers per second for CLIC-LHC based FEL-Pb collider are given on the last four columns of the Table 1. One can see that due to huge events rates, FEL-nucleus collider will give opportunity to determine main characteristics of excited nuclei in a very short time period comparing with traditional NRF methods.

## ACKNOWLEDGEMENTS

This work is supported by CERN, Turkish Atomic Energy Authority and Turkish State Planning Organizaton (DPT) under the Grants No 2003K120060, 2003K1201906-5 and 2002K120250.

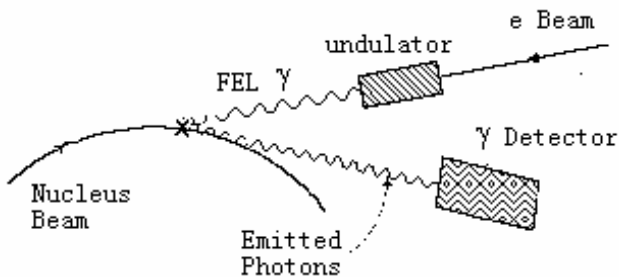

Fig. 1. The schematic view of FEL-Nucleus collider.

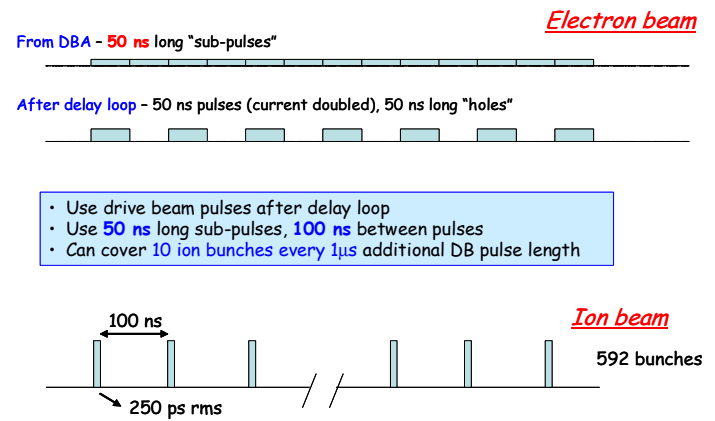

Figure 2. Matching the time structure of the drive beam and the LHC ion beam

Table 1. Main characteristics of some of the $^{208}$Pb nucleus excitatioe excitations.

| $E^*$ (MeV) | $\Gamma$ (eV) | $J^\pi$ | $\omega_{FEL}$ (keV) | $\sigma_{res}$ (cm$^2$) | $\sigma_{ave}$ (cm$^2$) | R/s |
|---|---|---|---|---|---|---|
| 4.0852 | 7.83·10$^{-1}$ | 2+ | 0.694 | 7.3·10$^{-22}$ | 1.39·10$^{-24}$ | 5.2·10$^5$ |
| 4.8422 | 99.72·10$^{-1}$ | 1- | 0.823 | 5.23·10$^{-22}$ | 1.07·10$^{-23}$ | 4.0·10$^6$ |
| 5.2926 | 13.16·10$^0$ | 1- | 0.899 | 2.62·10$^{-22}$ | 2.62·10$^{-24}$ | 2.4·10$^6$ |
| 5.5122 | 32.91·10$^0$ | 1- | 0.937 | 2.41·10$^{-22}$ | 0.14·10$^{-22}$ | 5.2·10$^7$ |
| 5.8461 | 11.54·10$^{-1}$ | 1+ | 0.993 | 21.5·10$^{-22}$ | 4.25·10$^{-24}$ | 1.5·10$^5$ |
| 5.9480 | 10.12·10$^{-1}$ | 1- | 1.011 | 2.07·10$^{-22}$ | 3.52·10$^{-25}$ | 1.3·10$^4$ |
| 6.2640 | 10.12·10$^{-1}$ | 1- | 1.064 | 1.87·10$^{-22}$ | 3.02·10$^{-25}$ | 1.1·10$^4$ |
| 6.3117 | 36.56·10$^{-1}$ | 1- | 1.072 | 1.06·10$^{-22}$ | 1.06·10$^{-24}$ | 3.9·10$^5$ |
| 6.3628 | 10.44·10$^{-1}$ | 1- | 1.081 | 1.81·10$^{-22}$ | 2.96·10$^{-25}$ | 1.1·10$^4$ |
| 6.7205 | 10.97·10$^0$ | 1- | 1.142 | 1.62·10$^{-22}$ | 2.64·10$^{-24}$ | 0.9·10$^5$ |
| 6.9800 | 50.64·10$^{-1}$ | - | 1.186 | - | - | - |
| 7.0635 | 28.61·10$^0$ | 1- | 1.200 | 1.47·10$^{-22}$ | 5.95·10$^{-24}$ | 2.2·10$^5$ |
| 7.0834 | 14.62·10$^0$ | 1- | 1.200 | 1.46·10$^{-22}$ | 3.62·10$^{-24}$ | 1.3·10$^5$ |
| 7.2430 | 15.67·10$^{-1}$ | - | 1.231 | - | - | - |
| 7.2780 | 15.67·10$^{-1}$ | - | 1.237 | - | - | - |
| 7.2789 | 14.00·10$^{-1}$ | 1+ | 1.237 | 1.38·10$^{-22}$ | 2.65·10$^{-25}$ | 0.9·10$^4$ |
| 7.3325 | 38.71·10$^{-0}$ | 1- | 1.246 | 1.36·10$^{-22}$ | 7.18·10$^{-24}$ | 2.7·10$^5$ |
| 7.6853 | - | - | 1.306 | - | - | - |
| 10.050 | - | - | 1.708 | - | - | - |
| 10.600 | - | - | 1.801 | - | - | - |
| 11.450 | - | - | 1.975 | - | - | - |